\documentclass[11pt,twoside]{article}

\usepackage{asp2006}
\usepackage{epsf}
\usepackage{psfig}
\usepackage{lscape}
\usepackage{amssymb}

\begin{document}

\title{Self-organization in Turbulent Molecular Clouds: Compressional versus Solenoidal Modes}   

\author{A. G. Kritsuk\altaffilmark{1}, S. D. Ustyugov\altaffilmark{2}, M. L. Norman\altaffilmark{1}, 
and P. Padoan\altaffilmark{1}}

\altaffiltext{1}{Physics Department and Center for Astrophysics \& Space Sciences, 
University of California, San Diego;   
9500 Gilman Drive, La Jolla, CA 92093-0424, USA}
\altaffiltext{2}{Keldysh Institute for Applied Mathematics, Russian Academy of Sciences,
Miusskaya Pl. 4, Moscow 125047, Russia}

\begin{abstract} 
We use three-dimensional numerical simulations to study self-organization in 
supersonic turbulence in molecular clouds. Our numerical experiments describe 
decaying and driven turbulent flows with an isothermal equation of state, sonic
Mach numbers from 2 to 10, and various degrees of magnetization. We focus
on properties of the velocity field and, specifically, on the level of its
potential (dilatational) component as a function of turbulent Mach number, 
magnetic field strength, and scale. We show how extreme choices of either 
purely solenoidal or purely potential forcing can reduce the extent of the 
inertial range in the context of periodic box models for molecular cloud 
turbulence. We suggest an optimized forcing to maximize the effective Reynolds
number in numerical models.
\end{abstract}

\section{Introduction}
Modern statistical theories of fragmentation of molecular clouds (MCs) and 
star formation are based on an interpretation of the non-thermal emission 
linewidths and their correlation with length scale in terms of supersonic 
turbulence (Kaplan \& Pronik 1953; Larson 1981; Heyer \& Brunt 2004). It is believed that both 
the star formation rate and the initial mass function of newly born stars 
are controlled by MC turbulence (e.g., Padoan et al. 2007; McKee \& Ostriker 2007; Padoan \& Nordlund 2009). 
There are competing views on the origin of this turbulence, which is either 
explained as a transient phenomenon associated with the cloud formation 
process or as if it were continuously driven by various energy sources (e.g., 
differential rotation, supernovae, stellar winds, protostellar outflows,
variable FUV background, etc. \citep{maclow04}). Since the typical Reynolds 
numbers in MCs are $\sim\!10^8$, expectations to find purely 
laminar regions in the cold star forming molecular gas on scales from 
$\sim50$~pc down to a few astronomical units should be pretty low \citep{elmegreen.04}.

The supersonic regime typical of MC turbulence is extremely hard to achieve 
in the laboratory and the information available from astronomical observations 
is limited \citep[e.g.,][]{heyer.04}. Most of what we know about the statistics 
of supersonic turbulence comes from large-scale numerical experiments intended 
to reproduce the basic non-linear processes operating in the energy 
cascade in the inertial range of scales \citep[e.g.,][]{kritsuk...07a}.
The effective Reynolds numbers normally achieved in such simulations are at 
most $\sim\!10^4$, i.e. much smaller than the realistic values. Since computational
resources are always limited, even with the most advanced and least dissipative 
numerical methods only a short stretch of the inertial interval can be 
captured at current grid resolutions up to $2048^3$ zones \citep{kritsuk...09a}.
Simulations also rely on a number of assumptions intended to simplify the
model and make it more tractable, such as the periodic boundary conditions 
and an isothermal equation of state. To achieve a better scale separation,
an approach of implicit large eddy simulations (ILES) is used \citep{sytine....00}. 
The effects of molecular viscosity are, thus, replaced by numerical diffusivity 
of purely artificial nature. In simulations involving magnetic fields, the 
magnetic diffusivity is also replaced by the effective one built into an ideal 
MHD solver in use. In these circumstances the effective magnetic Prandtl numbers 
usually achieved are on the order unity \citep[e.g.,][]{kritsuk...09b}.

The assumption of isothermality naturally restricts the physical box size to 
$L\lesssim5$~pc, as the presence of multiple thermal phases plays a role on larger 
scales. In these circumstances, an artificial stirring force is required to mimic 
the turbulent energy flux from larger-scale cascade that cannot be modeled 
directly due to a finite physical dimension of the computational domain. With some 
rare exceptions, most of the models rely on a purely solenoidal forcing 
on large scales to provide a better `boundary condition' for the inertial interval 
in the wavenumber space \citep[e.g.,][]{boldyrev..02a}. While the stellar energy 
sources would mostly generate compressive fluctuations, the choice of a solenoidal 
force was then justified by the small compressional-to-solenoidal ratio measured in 
simulations. 

With higher quality and larger simulations available today, we can now
reassess the domain of applicability of solenoidal forcing in isothermal simulations
of MC turbulence. To achieve this goal, we use various simulations we have performed in
the past to investigate self-organization in supersonic turbulence and quantify the
equilibrium compressional-to-solenoidal ratio in the inertial range as a function of 
the sonic and Alfv\'enic Mach numbers. We then come up with an optimized 
prescription for the large-scale forcing in isothermal periodic boxes that allows us
to achieve a better scale separation. We show that a poor choice of forcing
parameterization can easily lead to a complete elimination of the inertial range 
and result in rather ``pathological'' statistics, which have nothing to do with 
turbulence that fully develops far from boundaries and external forces at very high 
Reynolds numbers.

\section{Dilatational and solenoidal motions in supersonic turbulence} 

The velocity field ${\bf u}(x,t)$ can be decomposed into solenoidal and dilatational
parts ${\bf u}_s$ and  ${\bf u}_c$, such that ${\bf u}={\bf u}_s+{\bf u}_c$, 
${\bf \nabla\cdot u}_s=0$ and ${\bf \nabla\times u}_c=0$ via Helmholtz decomposition.
Let us consider $\chi(k)\equiv P({\bf u}_c,k)/P({\bf u}_s,k)$ as a measure of the flow
compressibility,\footnote{There are also alternative measures of compressibility, e.g. 
$\gamma\equiv\left<u_c^2\right>/\left<u_s^2\right>$, which gives the global ratio of the
specific kinetic energy contained in the dilatational and solenoidal modes; 
$\chi_c(k)\equiv P({\bf u}_c,k)/P({\bf u},k)$, which estimates the fraction of dilatational
modes in the velocity power spectrum as a function of $k$; 
and $r_{cs}\equiv\left<|{\bf \nabla\cdot u}|^2\right>/\left(\left<|{\bf \nabla\cdot u}|^2\right>+
\left<|{\bf \nabla\times u}|^2\right>\right)$, which represents the small-scale compressive
ratio. Both $\chi_c$ and $r_{cs}$ are bounded in the interval $[0,1]$, while $\chi$ and $\gamma$ 
can potentially take arbitrary positive values.} where $P({\bf a},k)$ is the three-dimensional 
power spectrum of a vector field {${\bf a}$}, and $k$ is the wavenumber. In an incompressible
fluid $\chi(k)\equiv0$, while in a compressible gaseous medium with purely potential (rotation-free)
velocity field $\chi(k)\equiv\infty$.

In fully developed supersonic isothermal turbulence, $\chi(k)$ describes a balance established 
via nonlinear exchange between the dilatational and solenoidal modes, which is ultimately 
controlled by the sonic ($M_s$) and Alfv\'enic ($M_A$) Mach numbers. In non-magnetized flows
($M_A=\infty$) at low sonic Mach numbers ($M_s\ll 1$), compressibility is very weak and 
$\chi(k)$ tends to settle at zero. At high turbulent Mach numbers ($M_s\gg 1$), the compressional-to-solenoidal
ratio hovers around $1:2$, which can be explained by simple geometrical considerations
\citep[e.g.,][]{nordlund.03}. In MHD turbulence, the natural tendency towards Alfv\'enization,
or dynamic alignment between the velocity field ${\bf u}$ and the divergence-free magnetic field
${\bf B}$ in the bulk of the volume away from shocks and dynamic rarefactions, 
results in suppression of dilatational activity. Thus, the presence of dynamically important 
magnetic fields would effectively reduce $\chi(k)$ in trans- and sub-Alfv\'enic turbulence 
\citep[e.g.,][]{boldyrev..02a}.

In simulations, besides $M_s$ and $M_A$, the ratio $\chi(k)$ would also depend on the content of 
solenoidal and dilatational modes in the large-scale forcing, $\chi_f$. If $\chi_f$ is far from 
the equilibrium ratio that corresponds to chosen values of $M_s$ and $M_A$, the effect of 
such forcing will be felt further down the hierarchy of scales and the inertial range in 
such simulations would shrink or disappear depending on what $\chi_f$ is enforced at the driving
scale $k_f$. Most of the simulations conservatively used a purely solenoidal 
forcing with $\chi_f=0$, with the exception of  $\chi_f\approx0.7$ in \cite{kritsuk...07a}.
\cite{schmidt..08} recently considered both $\chi_f=0$ and $\chi_f=\infty$ in $1024^3$ 
non-magnetized simulations at $M_s\approx5.5$ and found that the velocity scaling varies 
substantially with the large-scale forcing. \cite{federrath..08} also discovered that the 
density pdf in their compressively driven models does not bear a lognormal shape, see also 
\cite{schmidt....09}. \cite{federrath....09} showed that $\chi(k)\approx1.2$ at $k/k_{min}\in[3, 70]$ 
in their $1024^3$ simulation at $\chi_f=\infty$, while at $\chi_f=0$ they obtained the expected
$\chi(k)\approx0.5$ at $k/k_{min}\in[8, 30]$, where $k_{min}=2\pi$.\footnote{The same simulations also produced an unusual 
peak in $\chi(k)$ on small scales at $k/k_{min}\in[300, 400]$ with the peak values of $2.3$ and $4.0$ 
in the runs with solenoidal and compressive forcing, respectively.} 
\cite{schmidt09} found that a transition from $\chi_f=0$ to $\chi_f=\infty$ causes strong 
variations in spectral properties of turbulence in his large eddy simulations (LES). While 
these ``pathological'' 
statistics observed in simulations with purely compressive forcing clearly indicate 
a complete absence of an inertial range even at a grid resolution of $1024^3$ zones at 
$\chi_f=\infty$, they also hint at a possibility to optimize the problem setup by tuning the 
forcing to match the expected statistical equilibrium in the inertial range determined by the 
flow parameters. This would help to maximize the extent of the inertial range and thus to 
provide higher effective Reynolds numbers at the same computational cost.

\begin{figure}
\vspace{-0.5cm}
\centering
%
\plottwo{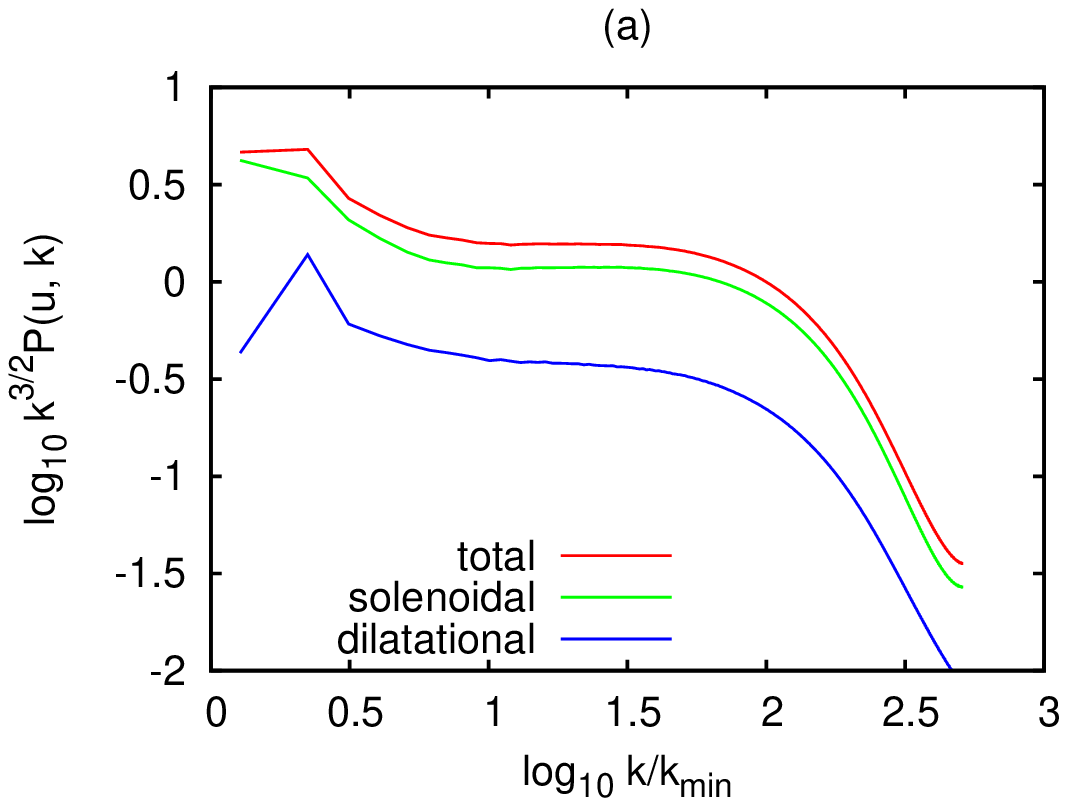}{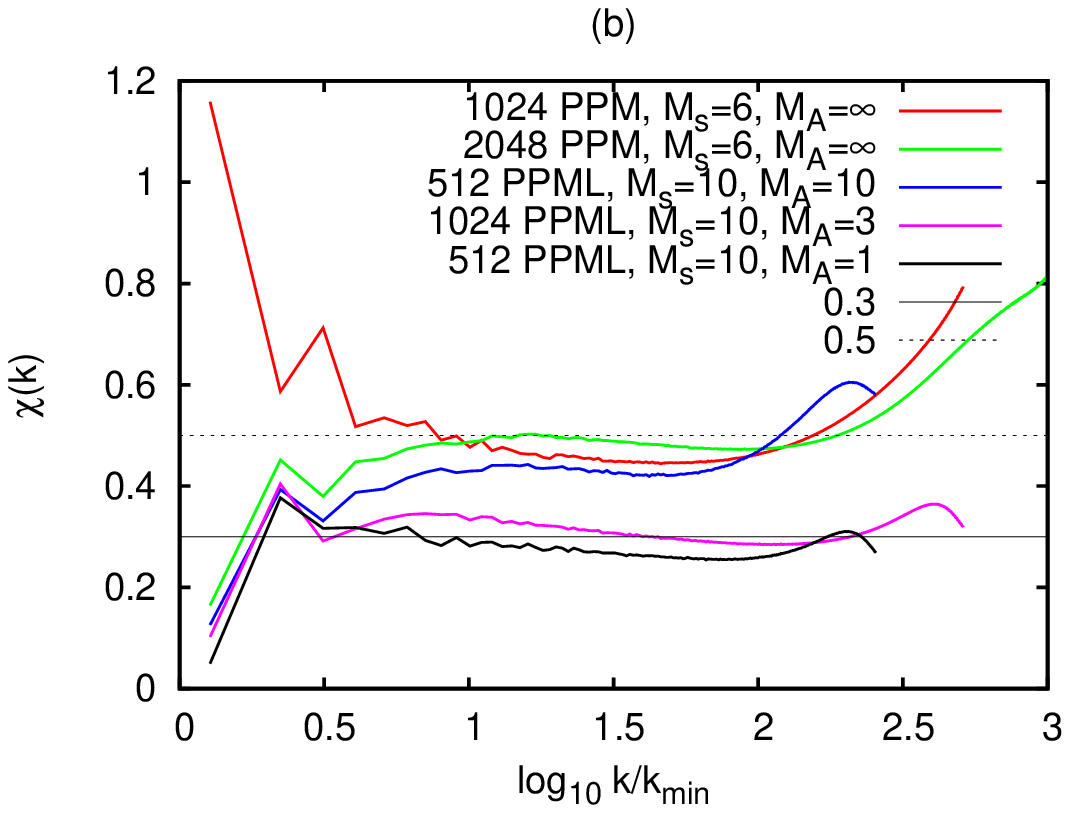}
\plottwo{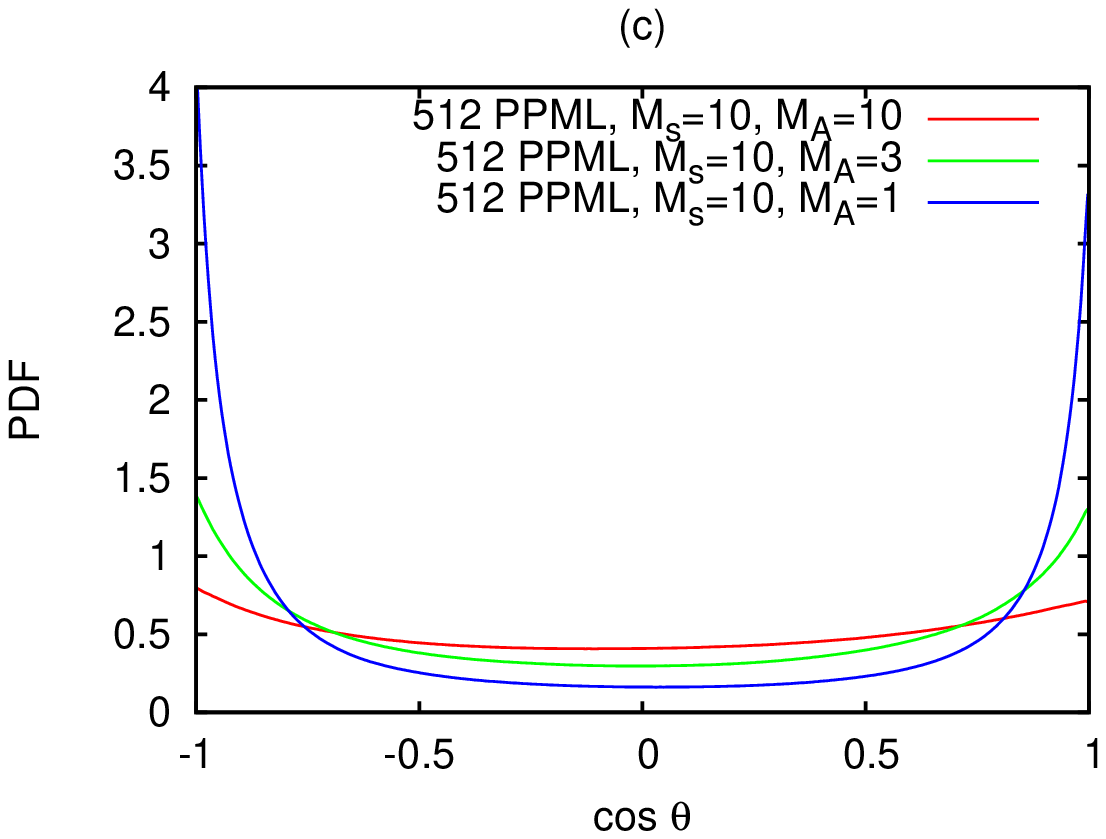}{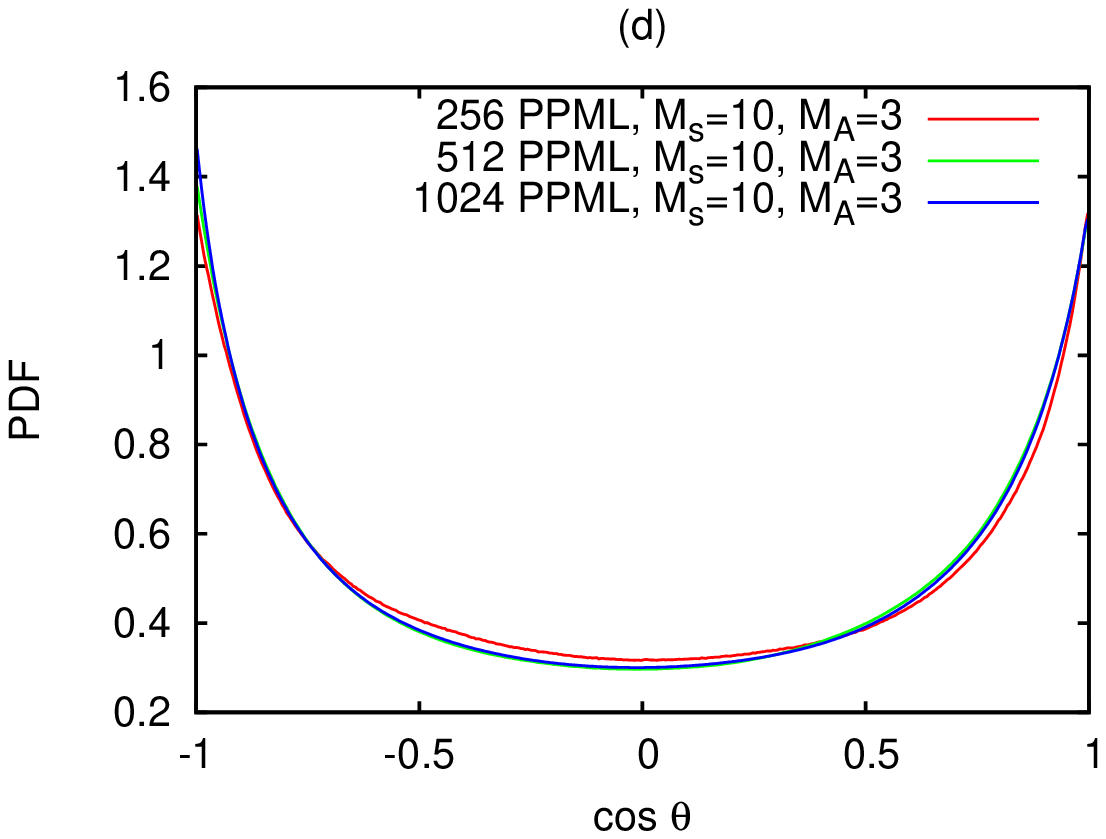}
\plottwo{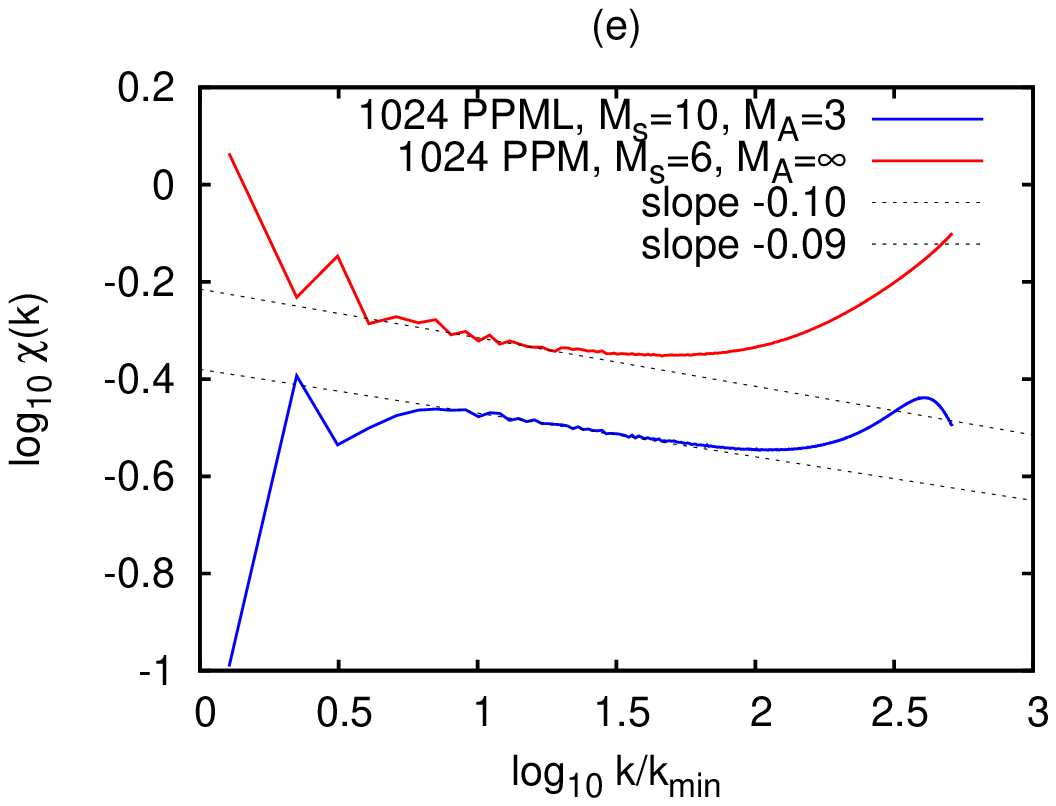}{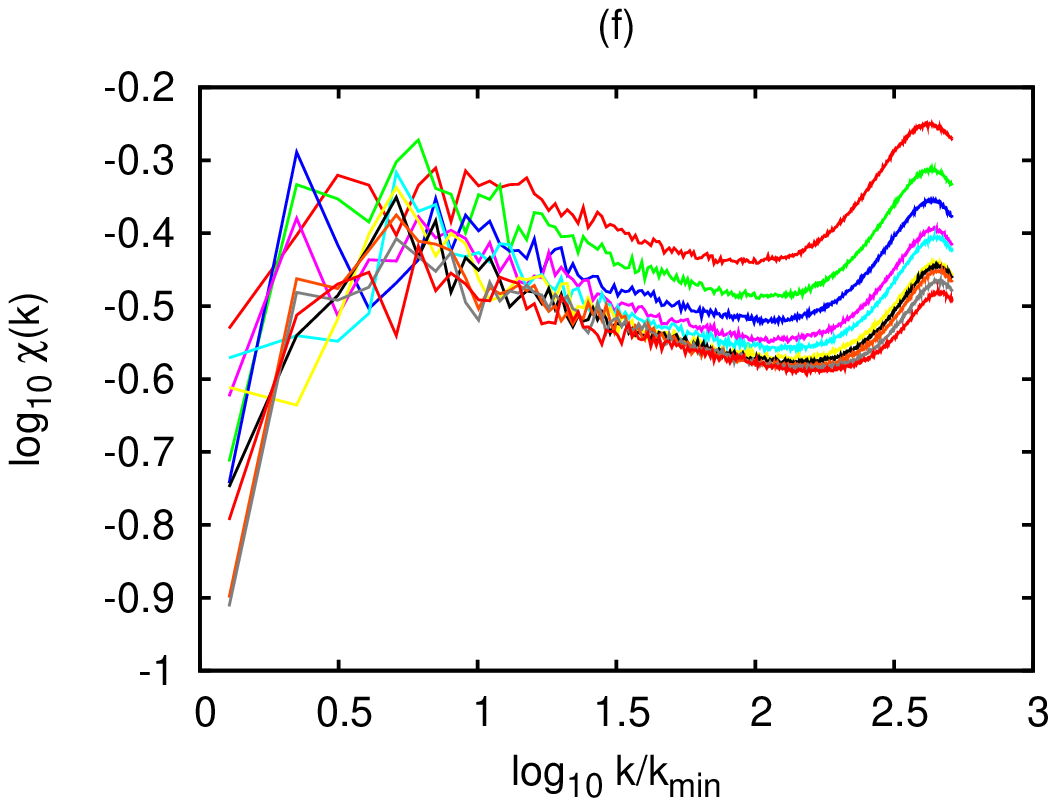}
\plottwo{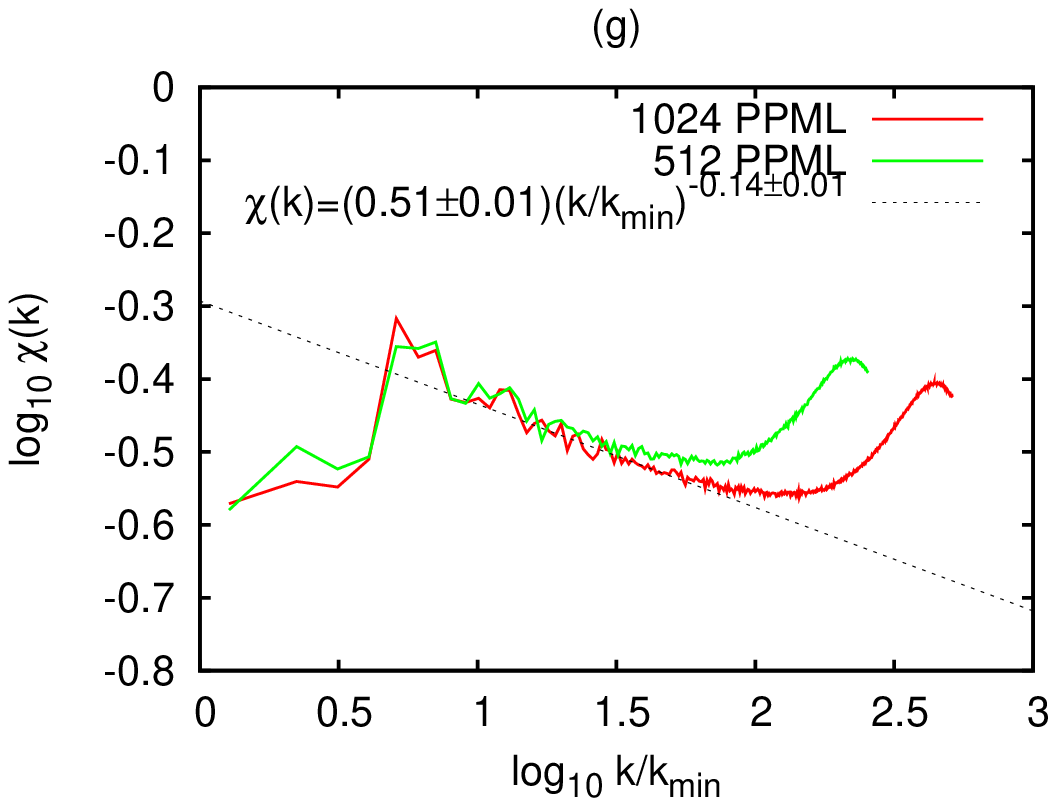}{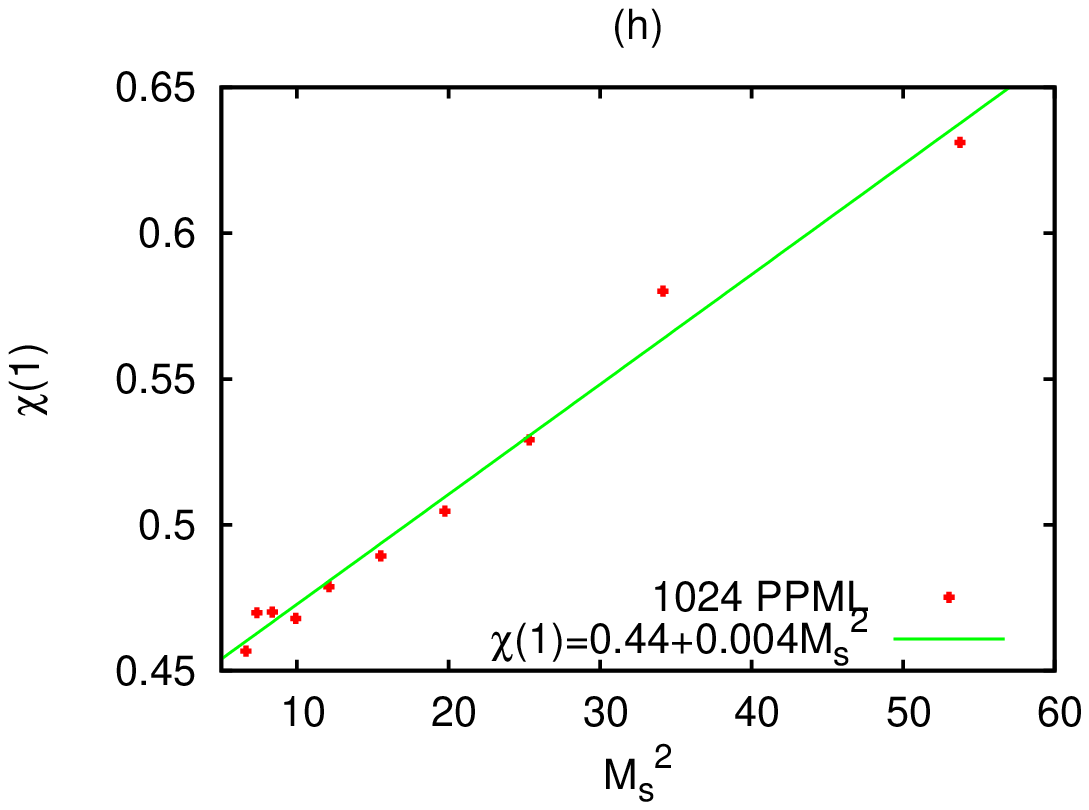}
\caption{\small Statistics of supersonic turbulence from simulations with PPM and PPML:
({\bf a}) velocity power spectrum, and spectra for dilatational and solenoidal parts, 
$M_s=10$ and $M_A=3$;
({\bf b}) $\chi(k)$ for simulations with
mixed forcing ($\chi_f=0.7$, red line) and with solenoidal forcing ($\chi_f=0$, all the rest);
({\bf c}) pdfs of the alignment angle for three $512^3$ simulations with $M_A=1$, 3, and 10;
({\bf d}) convergence of the $\cos{\theta}$ pdfs for runs at $256^3$, $512^3$, and $1024^3$;
({\bf e}) same as ({\bf b}), but for $log_{10}\chi(k)$ at $1024^3$ only;
({\bf f}) $\chi(k)$ for 10 snapshots from a $1024^3$ simulation of turbulence decay;
({\bf g}) convergence of $\chi(k)$ for the 5th snapshot from $512^3$ and $1024^3$ decay simulations;
({\bf h}) $\chi(k/k_{min}=1)$ as a function of $M_s^2$ from a $1024^3$ simulation of turbulence decay.
}
\vspace{-0.5cm}
\label{figone}
\end{figure}

To explore the effects of artificial large-scale forcing on the velocity field statistics at
scales adjacent to the forcing range, we collected data from various isothermal simulations 
with and without magnetic fields \citep{kritsuk...07a,kritsuk...09a,kritsuk...09b}. 
The nonmagnetized runs utilized the Piecewise Parabolic Method (PPM) of \citet{colella.84} 
implemented in the ENZO code.\footnote{\tt http://lca.ucsd.edu/projects/enzo} 
The MHD simulations were carried out with our Piecewise Parabolic Method on a Local Stencil 
\citep[PPML,][]{ustyugov...09}.

Figure 1{\bf a} gives an example of the velocity power spectrum $P({\bf u},k)$ and products
of Helmholtz decomposition for a $1024^3$ simulation with $\chi_f=0$, $M_s=10$ and $M_A=3$ 
\citep{kritsuk...09b}. One advantage of MHD simulations, even in the super-Alfv\'enic regime,
where the weak field is dynamically unimportant in most of the simulation domain,
is the absence of visible bottleneck contamination in the inertial subrange adjacent to the
dissipation range. This simplifies the discussion of the inertial range scaling for $\chi(k)$.
In Fig.~1{\bf b} we collect the $\chi(k)$ functions from various simulations. For instance,
the red and green curves represent two non-magnetized simulations with PPM at $M_s=6$ with $\chi_f=0.7$ 
\citep[$1024^3$, ][]{kritsuk...07a} and with $\chi_f=0$ \citep[$2048^3$,][]{kritsuk...09a}. In 
both cases $\chi(k)$ is close to the asymptotic value of 0.5 at wavenumbers $k/k_{min}\in[10, 100]$, 
as expected. Similar $\chi$-levels were also achieved in solenoidally driven non-magnetized 
simulations at $M_s\gtrsim5$ by others \citep{pavlovski..06,schmidt....09}. Consistently lower 
levels of $\chi\lesssim0.15$ were found in (adiabatic) simulations at $M_s\approx1$ 
\citep{pouquet..91,porter..94,porter...99,porter..02}. 
As the strength of magnetic field fluctuations climbs up to equipartition with turbulent kinetic 
energy in our sequence of PPML simulations with $M_A=10$, 3, and 1, the average level of 
$\chi$ drops from 0.5 to below 0.3 for the sonic Mach number fixed at $M_s=10$ 
\citep[see also][]{boldyrev..02a,boldyrev..02b}. 

To illustrate the effects of dynamic alignment in magnetized supersonic flows, in Fig.~1{\bf c} 
we show the probability density functions (pdfs) of the cosine of the alignment angle, 
$\cos{\theta}\equiv{\bf u\cdot B}/\sqrt{u^2B^2}$, for three $512^3$ PPML simulations
at $M_s=10$ and $M_A=10$, 3, and 1. In the most super-Alfv\'enic case at $M_A=10$,
the alignment is rather weak. It gets substantially stronger at $M_A=1$, when the 
equipartition of turbulent magnetic and kinetic energies is reached. Fig,~1{\bf d} 
shows that the pdf of the alignment angle is well converged already at resolution of 
$512^3$ in a series of representative numerical experiments with $M_A=3$ and grid 
resolutions of $256^3$, $512^3$, and $1024^3$.

Let us get back to Fig.~1{\bf b} and look at the shape of the compressional-to-solenoidal ratio
$\chi(k)$ in more detail. Since the two non-magnetized (PPM) simulations differ only in $\chi_f$ 
and in the grid resolution, time-average spectra in Fig.~1{\bf b} indeed capture the effects of 
large-scale forcing that operates at $k_f/k_{min}\in[1,2]$ in all cases shown. If one assumes that a 
small negative slope $\chi(k)\sim k^{-0.1}$ seen at $k/k_{min}\in[10,60]$ (which can be traced to 
some extent in all simulations presented in this figure, see also Fig.~1{\bf e}) is real, i.e. 
forms as a result of self-organization in supersonic turbulence, then it would seem that an 
isotropic forcing with $\chi_f\in[0.6,0.7]$ would be an optimal choice for $M_s\approx 6$. At the 
same time, the green line in Fig.~1{\bf b} shows that the effects of enforcing $\chi_f=0$ at 
$k_f/k_{min}\in[1,2]$ are felt at least up to 
$k=16k_{min}$, which is substantially larger than $k_f$. A similar decline towards smaller wavenumbers
is seen in the blue and pink curves, corresponding to $M_A=10$ and 3, respectively. The
trans-Alfv\'enic run at $M_A=1$ shown in black does not have that feature.

To better recognize the slope in $\chi(k)\sim k^{-0.1}$, we replot the two $1024^3$ results in a 
log-log plot, where it can be seen better (Fig.~1{\bf e}). Is this slope real? Does it depend on
the forcing? Is it related to specifics of numerical dissipation at small scales? To address these
questions, we explored results of an MHD simulation of decaying turbulence, where the effects of 
continuous driving are minimized. This $1024^3$ PPML simulation was carried out as part of KITP07
code comparison project.\footnote{\tt http://kitpstarformation07.wikispaces.com/Star+Formation+Test+Problems}
The simulation follows a free decay of turbulence from a developed statistical steady state with 
$M_s\approx10$ and $M_A\approx10$ down to $M_s\approx2$. Figure~1{\bf f} shows $\chi(k)$ for 10 
flow snapshots equally spaced in time. As can be seen, with no forcing, the slope of $-0.13\pm0.03$
is clearly present even though in this case we only have instantaneous power spectra and 
the data are rather noisy. In Fig.~1{\bf g} we show $\chi(k)$ obtained for the 5th snapshot
at $512^3$ and $1024^3$ to illustrate grid convergence with PPML. We also compared PPML results 
with those from three other popular numerical methods and implementations for compressible ideal MHD
\citep[ZEUS, FLASH, RAMSES, see][]{kritsuk.2009c} at a grid resolution of $512^3$ zones and found 
an excellent agreement.
It seems that the slope is indeed real and does not depend on forcing or numerical dissipation.
We derived a simple fitting formula for $\chi(k)$ in this decaying turbulence model, assuming
a fixed slope of $-0.13$ and fitting a normalization constant $\chi(k/k_{min}=1)$ against $M_s^2$, as 
shown in  Fig.~1{\bf h}. The approximation valid in the inertial range for rms turbulent Mach numbers 
$M_s\in[1,10]$, 
\begin{equation} 
 \chi(k)\approx\left(0.44+0.004M_s^2\right)(k/k_{min})^{-0.13},
\end{equation}
is not unique, but still can be used as a guide in future experiments with optimized forcing
for supersonic super-Alfv\'enic turbulence.

\section{Conclusion}
We explored the problem of forcing optimization to maximize the effective Reynolds numbers
achieved in simulations of supersonic molecular cloud turbulence. Our results demonstrate
that for low sonic Mach numbers and/or Alfv\'enic Mach numbers $M_A\lesssim1$ solenoidal 
forcing is reasonable. At the same time, for sonic and Alfv\'enic Mach numbers above 3,
solenoidal forcing is inappropriate, a proper mixture of dilatational and solenoidal modes 
is required for optimal representation of the inertial range. A purely compressional
driving is impractical as it does not allow to study the scaling properties of fully
developed supersonic turbulence.

\acknowledgements {\small

This research was supported in part by the National Science Foundation through 
grants AST-0607675, AST-0808184, and AST-0908740 as well as through TeraGrid resources 
provided by NICS, PSC, SDSC, and TACC (MCA07S014 and MCA98N020).} AK appreciates
discussions with Stanislav Boldyrev on dynamic alignment.


\end{document}